# Low-temperature ferroelectric phase and magnetoelectric coupling in the underdoped $La_2CuO_{4+x}$

Z. Viskadourakis[1], I. Radulov[1], A. P. Petrović[2], S. Mukherjee[1,3], B. M. Andersen[3], G. Jelbert[2,4], N.S. Headings[5], S.M. Hayden[5], K. Kiefer[6], S. Landsgesell[6], D.N. Argyriou[6,7] and C. Panagopoulos[1,2,4,8]

[1] Institute of Electronic Structure and Laser, Foundation for Research and Technology Hellas, Heraklion 70013, Greece

[2] Division of Physics and Applied Physics, Nanyang Technological University, 637371 Singapore

[3] Niels Bohr Institute, University of Copenhagen, DK-2100 Copenhagen, Denmark

[4] Cavendish Laboratory, University of Cambridge, Cambridge CB3 0HE, U.K.

[5] H. H. Wills Physics Laboratory, University of Bristol, Tyndall Ave., Bristol BS8 1TL, U.K.

[6] Helmholtz-Zentrum-Berlin für Materialien und Energie, D-14109 Berlin, Germany

[7] European Spallation Source ESS AB, P.O. Box 176, SE-221 00 Lund, Sweden

[8] Department of Physics, University of Crete, Heraklion 71003, Greece

We report the discovery of ferroelectricity below 4.5 K in highly underdoped $La_2CuO_{4+x}$ accompanied by slow charge dynamics which develop below T~40 K. An anisotropic magnetoelectric response has also been observed, indicating considerable spin-charge coupling in this lightly doped "parent" high temperature copper-oxide superconductor. The ferroelectric state is proposed to develop from polar nanoregions, in which spatial inversion symmetry is locally broken due to non-stoichiometric carrier doping.

## I. INTRODUCTION

Twenty five years have passed since the initial discovery of high-temperature (high-$T_c$) superconductivity in the cuprates. The "parent" compounds of this family constitute archetypal antiferromagnetic Mott insulators [1], which upon carrier-doping exhibit a wide range of novel ground states, including glassy magnetic phases and unconventional superconductivity [2, 3]. However, determining the nature of the charge correlations coexisting with magnetic order has remained elusive, particularly in the highly underdoped limit.

Throughout this period it has been tacitly assumed that these materials do not exhibit ferroelectricity, since both their crystal structure and magnetic order display spatial inversion symmetry. Furthermore, the presence of mobile carriers at the Fermi level upon doping would seem to be a significant impediment to charge localization and ordering. In spite of this apparent impasse, several theoretical models have predicted that a ferroelectric ground state could indeed develop in these materials [4, 5], a concept encouraged by an early ultrasound study in $YBa_2Cu_3O_{6+x}$ [6]. Therefore, $La_2CuO_{4+x}$, which is structurally the simplest high-$T_c$ cuprate was chosen here in order to carry out a detailed analysis of its electronic polarization and charge dynamics in the strongly-underdoped limit. The main goal of this analysis is to identify and study the emergent ground state of these materials when the first charge carriers are added to the parent compound.

In this paper we report the discovery of ferroelectricity in $La_2CuO_{4+x}$ as a direct consequence of carrier-doping the undoped parent compound $La_2CuO_4$. In addition, a magnetoelectric effect is also observed, indicating that the charge and magnetic orders are coupled. A theoretical scenario is provided, which coherently accounts for these experimental discoveries. The possible implications of our results for the high-$T_c$ cuprate phase diagram are also discussed within this scenario.

## II. EXPERIMENTAL DETAILS

Single crystals of $La_2CuO_{4+x}$ were grown in Bristol using the traveling solvent floating zone technique [7]. The as-grown crystals were annealed at 800 $^o$C for 48 h to set the oxygen stoichiometry. From the initial crystals, thin plates were cut with the thinnest dimensions either along the c-axis or in the ab-plane. The magnetization of the samples was measured using commercial Quantum Design MPMS-XL SQUID magnetometers, and their Néel transition was found to be $T_N$=320 K (fig. 1a), slightly lower than the maximum $T_N$ seen in this compound [1]. The room temperature carrier concentration of the samples was measured employing the Van der Pauw method. It was found to be exceptionally low at $n=10^{17}$ cm$^{-3}$, almost an order of magnitude smaller than the values reported earlier [8]. The excess oxygen level was also estimated to be 0.3±0.03 mole%. It is hence clear that the crystals under investigation do not exhibit metallic behavior at low temperatures, thus fulfilling the requirements for performing accurate dielectric spectroscopy and electrical polarization experiments.

The impedance and loss of the samples were measured using both an LCR meter and a capacitance bridge, over a wide frequency range (77 Hz – 2 MHz). We initially measured the electric polarization using a pyroelectric current method: the sample was poled by applying the desired electric field at T=50 K and subsequently cooling to 2 K. The electric field was then removed and both the pyroelectric current and the sample temperature were recorded as a function of time, while warming the sample at a constant heating rate of 3 K min$^{-1}$. For magnetic field-dependent polarization measurements, the magnetic field was applied before the electric field was turned off and the pyrocurrent then measured as described above. We extract the temperature-dependent polarization at a fixed magnetic field from a complete time-dependent pyrocurrent measurement. The magnetic field dependence of the polarization was evaluated by plotting the polarization at a chosen temperature for a range of applied magnetic fields. Multiple measurements were performed under the same experimental conditions to estimate the systematic error of the measurement, which was found to be less than 5%. For the electric measurements, silver paint contacts were placed onto the largest faces of the samples. The effect of the contacts on the pyroelectric current measurements was negligible. Measurements of the polarization were performed both in Heraklion and in Berlin using different experimental set ups, utilizing 6 T and 14 T magnets, respectively. The dielectric constant was measured both in Heraklion and in Cambridge, the magnetotransport in Heraklion and in Berlin, and the magnetization in Heraklion and in Singapore. The magnetotransport and magnetic experiments were performed on a number of crystals, also different electrical contacts and reproduced several times over a period of 28 months.

## III. RESULTS AND DISCUSSION

Figure 1b shows the real part of the out-of-plane dielectric constant $\varepsilon'_c$. For high frequencies (f>5kHz), $\varepsilon'_c(T)$ exhibits a step-like decrease, as the temperature is lowered. This feature shifts to higher temperatures as the frequency increases, a typical signature of dielectric relaxation process. For f<5kHz, $\varepsilon'_c$ shows a broad maximum between 30 K and 45 K, which shifts to higher temperatures and decreases with increasing frequency. A similar behavior has been reported for $La_2Cu_{1-x}Li_xO_4$ and $La_{2-x}Sr_xCuO_4$ single crystals [9]. The aforementioned behavior in $\varepsilon'_c$ may be effectively explained in the frame of dipolar relaxation from charge hopping. We note however, in the Li and Sr doped $La_2CuO_{4+x}$ materials this feature is believed to originate from the polarization of electronic domains with a range of characteristic resonant frequencies. A similar behavior has been observed in relaxor ferroelectric materials characterized by a diffuse phase transition and the freezing of short-range cluster-like ferroelectric order [10-11]. It corresponds to a relaxation process with an abrupt increase in $\varepsilon'_c(T)$ as indicated by the dashed line in Fig. 1b. Although we cannot rule out the contribution of charge hopping, as shown later the cluster-like ferroelectric model is more plausible because it explains both the permittivity behavior and the observed magnetoelectric effect.

Figure 2a shows that both the in-plane and out-of-plane polarization in $La_2CuO_{4+x}$ increase abruptly below 4.5 K, indicating the onset of ferroelectricity. Note that a spontaneous polarization develops regardless of whether the crystal is cooled in zero or non-zero electric field. The raw data yields a nearly isotropic polarization, with the electric field-cooled in-plane and out-of-plane values reaching $P_{ab}$=85 nC cm$^{-2}$ and $P_c$=75 nC cm$^{-2}$ respectively. The weak anisotropy is consistent with the behavior of both the conductivity and the dielectric constant [12, 13,], which tend isotropic when the Néel temperature of the sample is maximized (i.e., when the doping is minimal). $P_{ab}$ is also slightly greater than $P_c$ when the sample is cooled in zero electric field (inset of Fig. 2a). However, field-cooling has a larger effect on $P_c$ than $P_{ab}$, implying an enhanced polarizability along the c-axis and some disorder in the ab-plane. This is supported by the hysteresis loops depicted in Figs 2b and 2c. The large dielectric anomaly and the relaxor behavior in the charge dynamics indicate a dominant electronic contribution towards the polarization. In addition, the observation of non-zero polarizations along both the c-axis and ab-plane implies either that the spontaneous polarization is not aligned with a crystal axis, or that it arises from an ensemble of polar nanoregions with varying polarization vectors.

The presence of broken inversion symmetry in the $La_2CuO_{4+x}$ structure is an essential precondition for the formation of a short-range charge ordered ground state. In improper ferroelectrics, local breaking of spatial inversion is observed and/or predicted due to the presence of mechanisms including non-collinear magnetic ordering (such as a magnetic spiral structure) [14] and displacive structural transitions [15, 16]. Although we cannot completely discount the possibility of one of these mechanisms driving ferroelectricity in the cuprates, there is no experimental evidence for any of these instabilities at such low doping in these materials. Here, a more likely explanation for the ferroelectric cluster formation is due to the presence of the interstitial oxygen atoms, which is a natural consequence of carrier-doping $La_2CuO_{4+x}$ with excess oxygen. These dopant atoms occupy non-stoichiometric positions in the crystal lattice [17, 18] and lead to the formation of local electronic dipoles. Similar dipoles in other materials create the so called polar nanoregions (PNR) [10], which may interact to form a glass-like relaxor or even an ordered ferroelectric state, depending on the concentration and the polarizability of the host [19]. We note that although our experiments do not allow a quantitative estimate of the polar nanoregion size, the origin of the latter (due to electronic carrier doping) suggests a length-scale comparable to a few lattice spacings [20].

Charge fluctuations within the polar nanoregions slow down with decreasing temperature before freezing into a kinetic glass, creating frequency dispersions in the dielectric constant similar to those seen in Fig. 1b. Eventually, a stable electrical polarization appears, though often significantly below the freezing temperature [21]. The relaxor trends in the dielectric constant and the static ferroelectric order developing at low temperature are therefore strongly supportive of a polar nanoregion scenario.

Returning to the electric field-dependent polarization loops (Figs. 2b and 2c) we observe a hysteresis at T=2.5 K, which constitutes further evidence for relaxor ferroelectricity. These so-called "slim loops" result from a high-field orientational alignment of the polar nanoregions, which is mostly lost upon removal of the electric field (unlike the behavior of hysteresis loops formed by dipolar glass states) [21]. The small observed remnant polarization is thus indicative of short-range co-operative freezing of polar nanoregions. At higher temperatures, it becomes more difficult to detect the polar nanoregions, since their alignment is randomized and the global polarization averages to zero; furthermore, the conductivity of the crystal becomes sufficiently high hence impeding accurate pyroelectric measurements. However, ac experiments (Fig. 3) reveal that above 5 K the polarization tends asymptotically to zero, exhibiting a high temperature tail similar to relaxor ferroelectrics [11, 22, 23]. This persists up to at least 30 K (this value depends on the excitation frequency of the experimental probe), despite the clear absence of any stable polarization within this temperature range. We observe no charge signature around $T_N$ except for a small change in the gradient of the dielectric constant, which corroborates similar work in the literature [24].

This apparent independence of the antiferromagnetic ordering and charge dynamics does not however, preclude a spin-charge correlation in $La_2CuO_{4+x}$. Each $CuO_2$ plane also possesses a weak ferromagnetic moment aligned along the c-axis, originating from a spin canting due to Dzyaloshinskii-Moriya (DM) interactions [25]. In zero magnetic field, the direction of this moment is reversed between adjacent planes and hence the global moment is zero. However, above a critical field $H_c$ (applied along the c-axis) this moment is flipped on alternate planes, resulting in weak ferromagnetic order. Magnetoresistance measurements [13, 26] reveal a large jump at $H_c$, implying strong spin-charge correlations: it is therefore, prudent to investigate the magnetic field dependence of our observed polarization.

For H//c, $P_c$ initially rises by 30% and exhibits a broad maximum around 5 T (coinciding with the spin-flop transition seen in the c-axis magnetotransport), before falling (Fig. 4a). This implies that a ferromagnetic alignment of the $CuO_2$ spins does not favor electrical polarization. In contrast, for H//ab $P_c$ is suppressed, saturating at roughly 80% of its zero-field value above 3 T. Furthermore, Figs 4b and 4c illustrate that the magnetic field has no effect on the ferroelectric ordering temperature, which remains constant at 4.5 K. The magnetoelectric coupling therefore, appears to be rather weak in $La_2CuO_{4+x}$. Measurements of the ab-plane polarization show that $P_{ab}$ decreases with magnetic field in both H//c and H//ab orientations, albeit with a small kink near 5 T for H//c (Fig. 5d).

Since the magnetic space group of $La_2CuO_{4+x}$ in the antiferromagnetic region is represented by the centrosymmetric space group cmca, which forbids any linear magnetoelectric effect, our observations are likely to be caused by nonlinear couplings presumably generated by the dilute amount of dopants. Non-stoichiometric oxygen dopants break intra-unit cell inversion symmetry, locally destroying the centrosymmetric structure by creating polar distortions over correlation length-scales equivalent to the size of polar nanoregions. This also perturbs the magnetic order within the polar nanoregions via local distortions of the $CuO_2$ lattice, which frustrate the super-exchange mechanism responsible for antiferromagnetic ordering. Applying an external magnetic field will further modulate the magnetic order, provoking small displacements in the oxygen dopant positions via the DM interaction and thus explaining the observed magnetoelectricity. Such a scenario can allow for the presence of nonlinear magnetoelectric coupling terms and similar physics has been applied to explain the magnetoelectric behavior in other relaxor ferroelectric materials [27]. In fact using the above ideas we have been able to qualitatively explain the structure of the observed magnetoelectric curves [28]. Notably, unlike the case in many other multiferroic perovskites [29, 30], the DM interaction is not responsible for the onset of ferroelectricity in $La_2CuO_{4+x}$: it merely enables us to gently tune the emergent charge order. We note that although the very low concentration of dopants might suggest additional mechanisms for the emergent ferroelectric order, such as subtle non-centrosymmetric distortions [31], the observed relaxor ferroelectric state and associated magnetoelectric effect is more likely to be caused by the added charge carriers to the parent Mott insulator since they can provide a natural mechanism due to the presence of polarized regions with multiple relaxation time scales.

## IV. SUMMARY

In summary, we report evidence for ferroelectricity and its associated magnetoelectric effect in lightly charge-carrier doped $La_2CuO_{4+x}$. It follows naturally to question how the ferroelectric ground state evolves with increasing carrier doping towards the superconducting dome. For example, raising the oxygen content in $La_2CuO_{4+x}$ should increase the density of the polar nanoregions, since we are moving further away from stoichiometry. This will lead to a stronger dipolar exchange coupling between the nanoregions, which should enhance ferroelectricity. However, any such co-operative enhancement must be balanced against the increased conductivity of the sample due to the higher carrier density: not only would this render the experimental detection of the ferroelectric phase extremely difficult, but mobile charge carriers will also migrate to cancel out any electric dipoles. Developing an accurate method of probing slow charge dynamics at higher dopings is therefore an urgent yet non-trivial task.

## ACKNOWLEDGMENTS

We acknowledge the financial support by the European Union through MEXT-CT-2006-039047 and EURYI research grants. The work in Singapore was supported by The National Research Foundation.

## APPENDIX

### 1. Extrinsic factors in dielectric permittivity

Figure 1b depicts the temperature dependence of the out-of-plane dielectric permittivity $\varepsilon'_C$ measured at different frequencies. $\varepsilon'_C$ (T) shows a step-like decrease down to ~50 K, shifting to higher temperatures as the frequency increases. This feature is characteristic of a dielectric relaxation process. We also find that $\varepsilon'_C$ peaks just below 50 K and the height of the peak increases, and shifts to lower temperatures with decreasing frequency. This behavior is typical of relaxor ferroelectrics, characterized by a diffuse phase transition and the freezing in short-range cluster-like order [10, 11]. However, similar features may also be observed due to a contribution from the electrical contacts. It is therefore, important to clarify whether our observations are intrinsic. In response, we performed measurements on the same sample using two different types of contacts. In the first case, silver paint contacts were placed on the parallel surfaces of a sample. The contacts were left at room temperature to cure for 24 h. Platinum wires were used to connect the sample to the measurement probe. In the second case, the sample was mechanically pressed between two gold-coated plates of a capacitor. In both cases, the sample was measured using the same experimental parameters.

Experimental results for $\varepsilon'_C$ are shown in Fig. 5a. Above ~50 K the permittivity differs, suggesting a contribution from extrinsic effects. In addition, high vales of $\varepsilon'_C$ could be due to Maxwell-Wagner contributions from the depletion layers at the interface between the sample and the contacts [32]. We find however, in the temperature range around the dielectric peak both measurements give almost identical results, for all measured frequencies. Moreover, both measurements agree down to the lowest temperature measured. In addition, we measured two samples with different contact areas to check the contribution of the depletion layer area. The results are depicted in Fig. 5b. Notably, the peak in the dielectric permittivity is not affected, indicating that it does not arise from the depletion layer at the contacts. The abovementioned tests confirm again the intrinsic nature of the observed relaxation at temperatures near and below the peak of the dielectric constant. We cannot exclude the possibility however, that the contacts may contribute to the deviation of the curves at higher temperatures.

An alternative approach to investigate whether the permittivity dispersion originates from Maxwell-Wagner relaxation effects is reported by Wang *et al* [33]. The relaxor-like behavior caused by Maxwell-Wagner relaxation can be characterized by the temperature dependent peak height as described by

$$\varepsilon' = A_1/[B_1 + C_1 exp(-T_o/T)] \quad (A1)$$

and an Arrhenius-like relation

$$f \sim exp\left(-T_o/T\right) \quad (A2)$$

describing the permittivity peak position.

Figure 6 shows that both the peak height (Fig. 6a) and the peak position (Fig. 6b) deviate from the abovementioned fits, adding credence to the fact that the dielectric peak reflects the intrinsic properties of the material. Furthermore, the peak position of the permittivity may be fitted to a Vogel-Fulcher law however, the fitting parameters ($T_f$ = -68 K, E = 2200 K) suggest that the dipoles do not freeze. In fact, the peak-height values are better described by a Curie-Weiss law. (The purpose for using the Curie-Weiss fit is merely to obtain an estimate of the temperature of the peak in the dielectric permittivity as the frequency approaches zero.) As we show in Fig. 1b of the manuscript, in the low temperature regime all the curves collapse on one another, indicating freezing of the polar nanoregions (PNRs). Most notably the dielectric loss (tan ($\delta$)) <<1 at least below 10K where the ferroelectric order is observed, adding further credence to the intrinsic nature of the measured low temperature polarization, which is the focus of this work.

We note that charge-hopping does not affect the main results of this work i.e., the low temperature ferroelectric phase and magnetoelectricity in $La_2CuO_{4+x}$. Also, charge hopping does not preclude the inhomogeneous distribution of the charge carriers within the sample and does not contradict the PNR scenario, because in both cases the existence of charge dipoles is needed. In fact, PNRs are regions within which symmetry is locally broken, due to the presence of excess charge. Therefore, at higher temperatures, both PNR and dipole relaxations can coexist in $La_2CuO_{4+x}$.

## 2. Phenomenological description of the permittivity peaks

In order to support the intrinsic diffuse phase transition character of the permittivity peaks we applied the empirical model proposed by Santos *et al.* [34] for the description of the diffuse phase transition in ferroelectrics. We fitted our low frequency permittivity data against the proposed formula

$$\varepsilon' = \varepsilon'_{max}/1 + \left((T - T_{max})/\Delta\right)^{\xi} \quad (A3)$$

where $\varepsilon'_{max}$ is the permittivity peak value, $T_{max}$ the temperature in which the peak in the permittivity is observed and $\Delta$ is related to the peak broadening. Here $\xi=1$ indicates a 'normal' ferroelectric phase transition described by the Landau–Devonshire theory for ferroelectric phase transitions (first or second order) and $\xi=2$ the so-called 'complete' Diffuse Phase Transition (DPT) [35]. On the other hand, $\xi$ between these limits, i.e. 1 and 2 indicates a so-called 'incomplete' DPT, where the interaction between ferroelectric clusters is taken into consideration. Figure 7 shows good agreement between the experimental data and eq. (A3) with $\xi \sim 2$ indicating the diffuse phase transition character of the peaks, characteristic of relaxor ferroelectrics. The model fails to describe the high frequency permittivity data where dipole relaxation phenomena are expected to be enhanced.

## 3. Polarization data processing

Figure 8a shows the in-plane electric polarization loops at 2 K, 4 K and 5 K. Both the highest $P_{ab}$ values and the remnant polarization are suppressed with increasing temperature (similar to the behavior observed in the out-of-plane polarization). Figure 8b shows the in-plane polarization as a function of temperature, illustrating that due to the presence of a spontaneous polarization the measured polarization cannot be reversed by reversing the electric field. However, subtracting the data measured for E = 0 from the corresponding data measured for E(+) = 2 kV cm$^{-1}$, we obtain the black curve in Fig. 8c. Similarly, the red curve depicts the results obtained from the subtraction of the data obtained for E

= 0 from the data measured at E(-) = - 2 kV cm$^{-1}$. The two curves are now found to be almost symmetric, consistent with the behavior in other ferroelectric materials. The data shown in Fig. 8a were acquired using P(T, E) plots similar to Figs 8b and 8c.

### 4. Ac polarization measurements

To explore the ferroelectric character of La$_2$CuO$_{4+x}$ in greater depth, we performed ac polarization experiments as a function of temperature and applied electric field using a TF Analyzer 2000E Hysteresis Loop Tracer. The following procedure was adopted: the sample was first cooled to the lowest temperature in zero applied electric field. An initial triangular excitation pulse was applied to establish a polarization followed by three consecutive excitation pulses separated by a relaxation time of 1 s. During each pulse, the current was measured. Both voltage and current as a function of time are presented in Fig. 9a. Further to the observed capacitive features (almost constant current values upon charging and discharging the sample), the most interesting characteristic is the small peak in the current occurring just before the voltage is maximum (indicated by a broken line running through in Figs 9b and 9c). This behavior is typical of a material exhibiting ferroelectricity (in the case of non ferroelectric materials this peak, if observed should coincide with the peak in voltage). The polarization is determined by integrating the current with respect to time, giving a P-E hysteresis loop similar to the data depicted in the inset of Fig. 3.

### 5. Magnetic field dependence of the electrical resistivity

Figure 10 shows the magnetic field dependence of the normalized out-of-plane resistivity determined from measurements of the electric impedance at different temperatures. The measurement frequency for all the data shown here is 316 Hz. For H//c we observe a first order phase transition and a corresponding hysteresis at around 5.5 T and 3.5 T for T=70 K and 240 K, respectively due to the weak ferromagnetic transition associated with the DM interaction. In $La_2CuO_{4+x}$, the crystal anisotropy and the DM interaction fix the easy axis for the spins to the longer of the two in-plane orthorhombic directions (the b-axis). The direction of the weak ferromagnetic moments **L** induced by the DM interaction is fixed by the cross product **L**=**D**×**n$_0$** between the DM vector **D** (oriented along the shorter of the two in-plane orthorhombic directions – the a-axis), and the antiferromagnetic order parameter **n$_0$** (pointing along the b-axis), so that **L** is oriented along the c-axis, perpendicular to the $CuO_2$ planes of the crystal structure. A sufficiently large magnetic field applied along the c-axis can overcome the inter-plane antiferromagnetic coupling and induce a discontinuous spin-flop reorientation, causing the so-called weak ferromagnetic (first order) phase transition. Similar behavior has been reported previously [13]. The critical field is reduced at high temperatures following the decrease in **L** due to thermal fluctuations in **n$_0$**(T) [36]. For H//ab the magnetoresistance varies smoothly because the weak ferromagnetic moments induce a continuous rotation of **n$_0$** in the bc-plane [13, 36, 37] - our samples are twinned in the ab-plane and we therefore observe a spatially-averaged response - Fig. 10 (inset). Similar results were obtained for the in-plane resistivity.

## 6. References

1. B. Keimer, A. Aharony, A. Auerbach, R. J. Birgeneau, A. Cassanho, Y. Endoh, R. W. Erwin, M. A. Kastner and G. Shirane, Phys. Rev. B **45**, 7430 (1992).

2. A. Aharony, R. J. Birgeneau, A. Coniglio, M. A. Kastner and H. E. Stanley, Phys. Rev. Lett. **60**, 1330 (1988).

3. P. M. Grant, S. S. P. Parkin, V. Y. Lee, E. M. Engler, M. L. Ramirez, J. E. Vazquez, G. Lim, R. D. Jacowitz, and R. L. Greene, Phys. Rev. Lett. **58**, 2482 (1987).

4. R. A. Bari, Phys. Rev. B **7**, 2128 (1973).

5. J. L. Birman and M. Weger, Phys. Rev B **64**, 174503 (2001).

6.  V. Müller, C. Hucho, K. de Groot, D. Winau, D. Maurer, and K.H. Rieder, Solid State Commun. **72**, 997 (1989).

7.  S. Komiya, Y. Ando, X. F. Sun and A. N. Lavrov, Phys. Rev B **65**, 214535 (2002).

8.  N. W. Preyer, R. J. Birgeneau, C. Y. Chen, D. R. Gabbe, H. P. Jenssen, M. A. Kastner, P. J. Picone, and T. Thio, Phys. Rev. B **39**, 11563 (1989).

9.  G. R. Jelbert, T. Sasagawa, J. D. Fletcher, T. Park, J. D. Thompson, and C. Panagopoulos, Phys. Rev. B **78**, 132513 (2008).

10. L. E. Cross, Ferroelectrics **76**, 241 (1987).

11. A. Levstik, Z. Kutnjak, C. Filipic and R. Pirc, Phys. Rev. B **57**, 11204 (1998).

12. T. Thio, C. Y. Chen, B. S. Freer, D. R. Gabbe, H. P. Jenssen, M. A. Kastner, P. J. Picone, N. W. Preyer and R. J. Birgeneau, Phys. Rev. B **41**, 231 (1990)

13. C. Y. Chen, R. J. Birgeneau, M. A. Kastner, N. W. Preyer and T. Thio, Phys. Rev. B **43**, 392 (1991).

14. M. Kenzelmann, A. B. Harris, S. Jonas, C. Broholm, J. Schefer, S. B. Kim, C. L. Zhang, S.-W. Cheong, O. P. Vajk, and J. W. Lynn, Phys. Rev. Lett. **95**, 087206 (2005).

15. A. P. Levanyuk and D. G.Sannikov, Sov. Phys. Usp. **17**, 199 (1974)

16. T. Katsufuji, S. Mori, M. Masaki, Y. Moritomo, N. Yamamoto, and H. Takagi, Phys. Rev. B **64**, 104419 (2001).

17. F. Cordero, C. Grandini and R. Cantelli, Physica **C 305**, 251 (1998).

18. J. D. Perkins, J. M. Graybeal, M. A. Kastner, R. J. Birgeneau, J. P. Falck, and M. Greven, Phys. Rev. Lett. **71**, 1621 (1993).

19. G. Xu, Z. Zhong, Y. Bing, Z-G. Ye and G. Shirane, Nature Mater. **5**, 134 (2006).

20. M. A. Kastner, R. J. Birgeneau, C. Y. Chen, Y. M. Chiang, D. R. Gabbe, H. P. Jenssen, T. Junk, C. J. Peters, P. J. Picone, T. Thio, T. R. Thurston, and H. L. Tuller, Phys. Rev. B **37**, 111 (1988).

21. G. A. Samara, J. Phys. Condens. Matter. **15**, R367 (2003).

22. J. Hemberger, P. Lunkenheimer, R. Fichtl, H.-A. Krug von Nidda, V. Tsurkan and A. Loidl, Nature **434**, 364 (2005).

23. D. Viehland, J. F. Li, S. J. Jang, L. E. Cross and M. Wuttig, Phys. Rev. B **46**, 8013 (1992).

24. G. Cao, J. W. O'Reilly, J. E. Crow and L. R. Testardi, Phys. Rev. B **47**, 11510 (1993).

25. I. Dzyaloshinskii, J. Phys. Chem. Solids **4**, 241 (1958); T. Moriya, Phys. Rev. **120**, 91 (1960).

26. T. Thio, T. R. Thurston, N. W. Preyer, P. J. Picone, M. A. Kastner, H. P. Jenssen, D. R. Gabbe, C. Y. Chen, R. J. Birgeneau and A. Aharony, Phys. Rev. B **38**, 905 (1988).

27. V. V. Shvartsman, S. Bedanta, P. Borisov, W. Kleemann, A. Tkach and P. M. Vilarinho, Phys. Rev. Lett. **101**, 165704 (2008).

28. S. Mukherjee, B. M. Andersen, Z. Viskadourakis, I. Radulov, and C. Panagopoulos, Phys. Rev B. **85**, R140405 (2012).

29. I. A. Sergienko and E. Dagotto, Phys. Rev B **73**, 094434 (2006).

30. S-W. Cheong and M. Mostovoy, Nature Mater. **6**¸13 (2007).

31. M. Reehuis, C. Ulrich, K. Prokeš, A. Gozar, G. Blumberg, S. Komiya, Y. Ando, P. Pattison, and B. Keimer, Phys. Rev. B **73**, 144513 (2006).

32. P. Lunkenheimer, V. Bobnar, A. V. Pronin, A. I. Ritus, A. A. Volkov, and A. Loidl, Phys. Rev. B **66**, 052105 (2002).

33. C. C. Wang and S. X. Dou, Solid State Communications **149**, 2017 (2009).

34. I. A. Santos J. Phys.: Condens. Matter. **13**, 11733 (2001),

35. V. V. Kirilov, V. A. Isupov, Ferroelectrics **5,** 3 1973

36. L. Benfatto and M. B. Silva Neto, Phys. Rev. B **74**, 024415 (2006).

37. S. Ono, S. Komiya, A. N. Lavrov, Y. Ando, F. F. Balakirev, J. B. Betts, and G. S. Boebinger, Phys. Rev. B **70**, 184527 (2004).

**Figure Captions**

**Fig. 1. a.** Bulk magnetization data measured with the applied magnetic field along the c-axis. **b.** Temperature dependence of the out-of-plane dielectric constant $\varepsilon'_c$ for several frequencies. The black broken line indicates the high temperature envelope curve of the peaks, calculated assuming a Curie-Weiss law $C/(T-T_C)$, where C and $T_C$ are the Curie constant and temperature, respectively.

**Fig. 2. a.** Temperature dependence of both in-plane and out-of-plane polarization, $P_{ab}$ and $P_c$, respectively, for the maximum electric field applied during cooling. The inset shows the spontaneous polarization obtained when the sample is cooled in zero electric field. **b.** and **c.**: $P_{ab}$ and $P_c$ slim loops at T = 2.5 K. Loops are normalized with respect to the spontaneous polarization. Red and blue lines are guides to the eye.

**Fig. 3**. The solid blue line corresponds to the dc polarization obtained from measurements of the pyroelectric current. Black and red open circles correspond to ac measurements performed at 500 Hz and 5 kHz respectively, using a 2000 TF analyzer. Black and red broken lines are guides to the eye. Inset: Hysteresis loop measured at 500 Hz (red line) compared to the corresponding dc data (open circles; the blue line is a guide to the eye).

**Fig. 4. a.** Polarization at T = 2.5 K as a function of applied magnetic field for H//c and H//ab. Red and green broken lines are guides to the eye. **b.** and **c.**: Temperature dependence of $P_c$ with respect to the applied magnetic field. **d.** $P_{ab}$ at 2.5 K with respect to magnetic field for H//c and H//ab. Red and green broken lines are guides to the eye. **e.** and **f.**: Temperature dependence of $P_{ab}$ for several applied magnetic fields.

**Fig. 5. a.** $\varepsilon'_C(T)$ of $La_2CuO_{4+x}$ single crystal for 77 Hz and 1 kHz, measured using silver paint (black and green points) and gold-coated (red and blue lines) plates, respectively. **b.** Data for a measurement performed at 316 Hz for two samples with varying geometry. Similar results were obtained for all measured frequencies.

**Fig. 6. a**. Temperature dependence of the dielectric peak height fitted to eq. (A1) (red dashed line) and a Curie-Weiss law (blue dashed line). **b**. Temperature dependence of the dielectric peak position fitted to the Vogel-Fulcher law (blue dashed line) and the Arrhenius-like relations (eq. A2).

**Fig. 7:** Fits of the low-frequency permittivity curves to eq. (A3). Solid circles indicate experimental data, and broken lines the theoretical fits.

**Fig 8. a.** Electric field dependence of the in-plane polarization at different temperatures (broken lines are guides to the eye). **b.** In-plane polarization as a function of temperature for maximum, minimum and zero electric field. **c.** Subtracting the zero field measurement from the maximum (minimum) field yields the symmetric black (red) curves.

**Fig. 9**. **a**. Voltage and current vs. time data measured using a TF Analyzer 2000E Hysteresis Loop Tracer. **b.** and **c**. Current peak is observed before the voltage maxima indicating ferroelectricity.

**Fig. 10.** Normalized out-of-plane resistivity obtained from measurements of the electric impedance at f=316 Hz as a function of applied magnetic field (H//c) at fixed temperatures. A step-like increase is observed due to the spin-flop transition. The step occurs at lower fields as the temperature increases. The inset depicts $\Delta\rho_c/\rho_c$ vs. $H^2$ for H//ab at T=70 K.

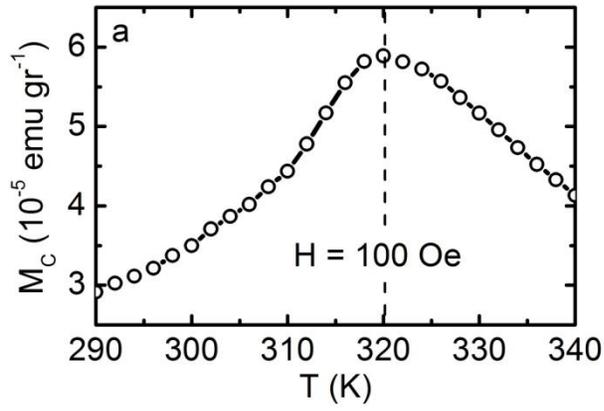
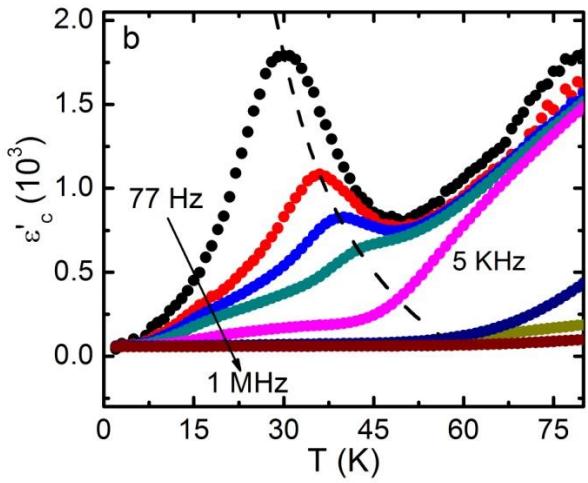

**Fig. 1**

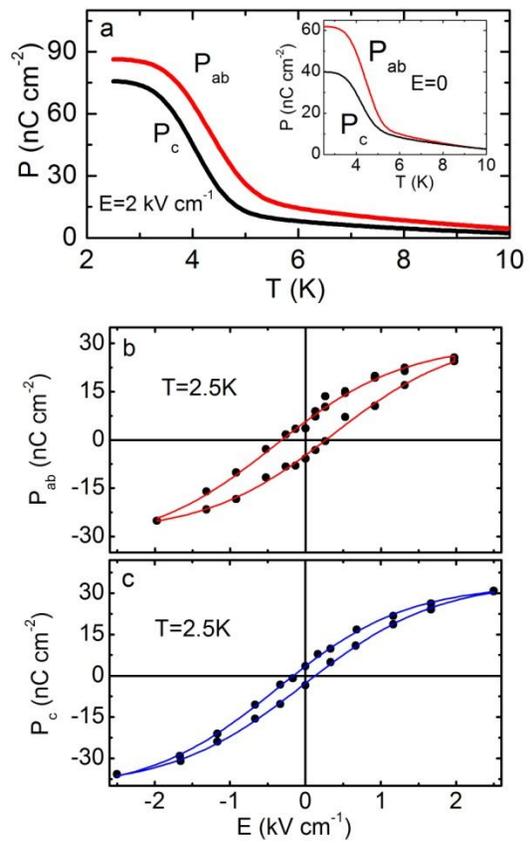

**Fig. 2**

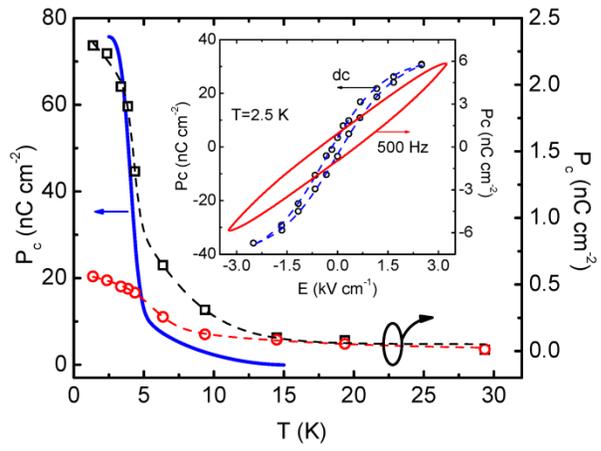

**Fig. 3**

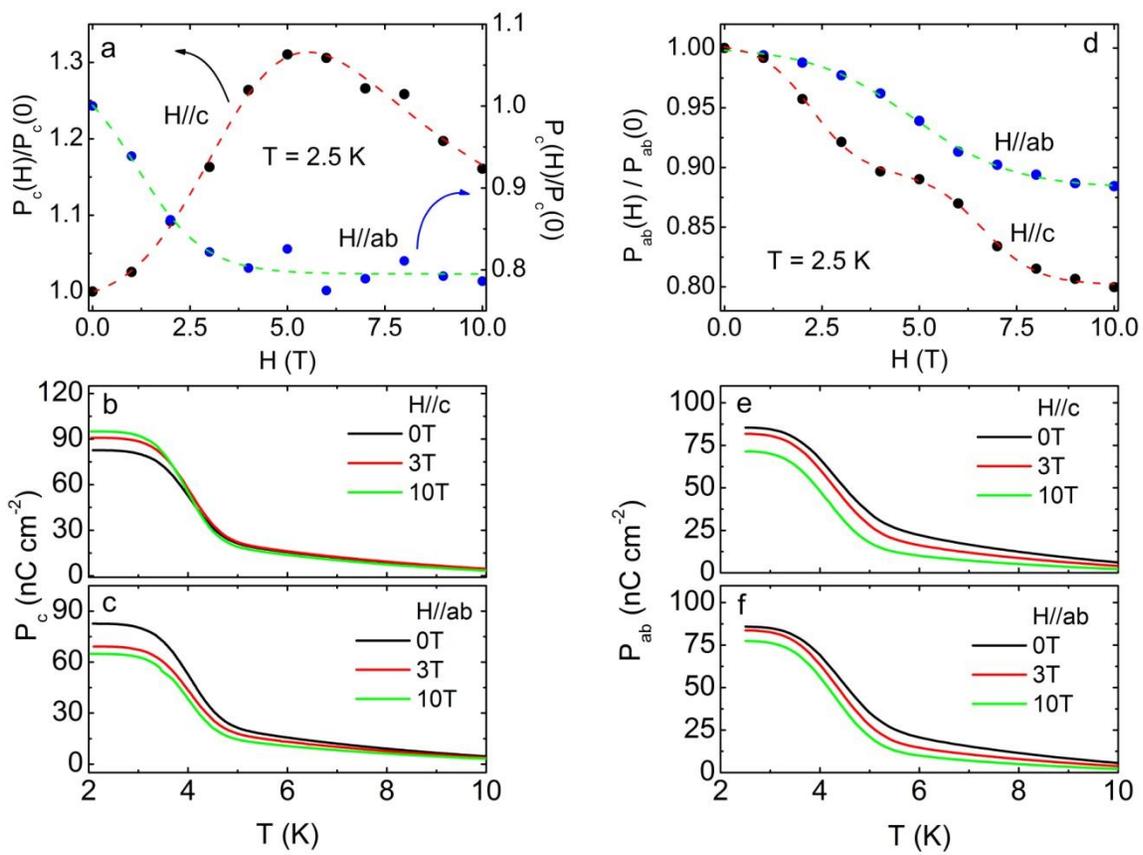

**Fig. 4**

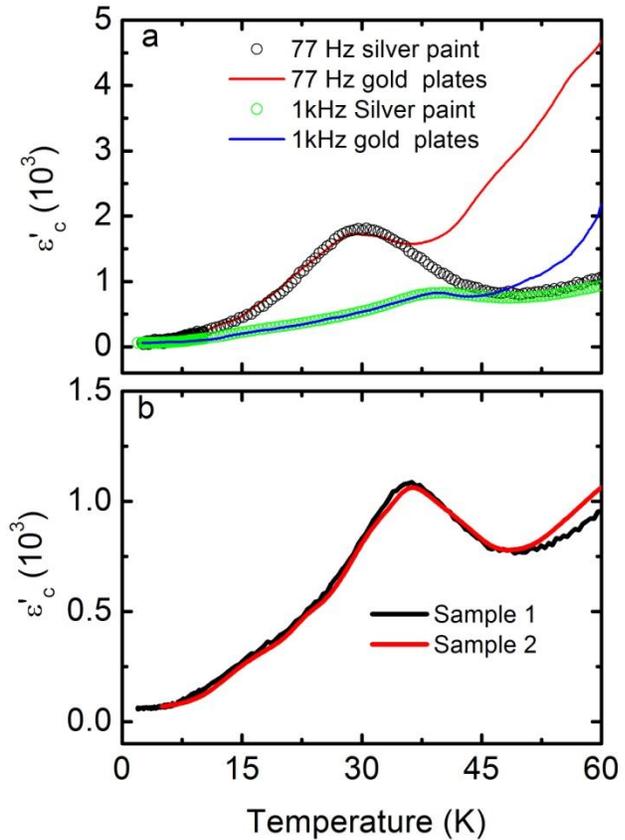

**Fig. 5**

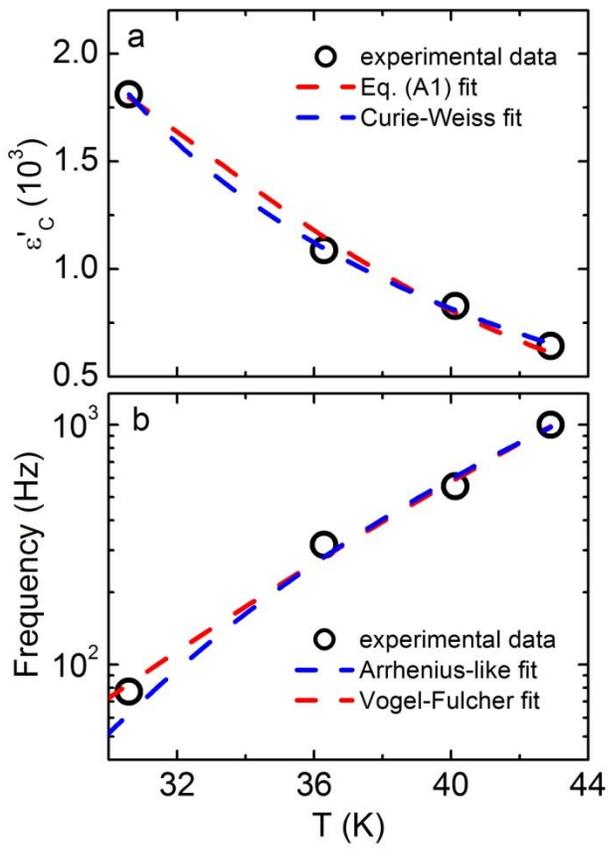

**Fig. 6**

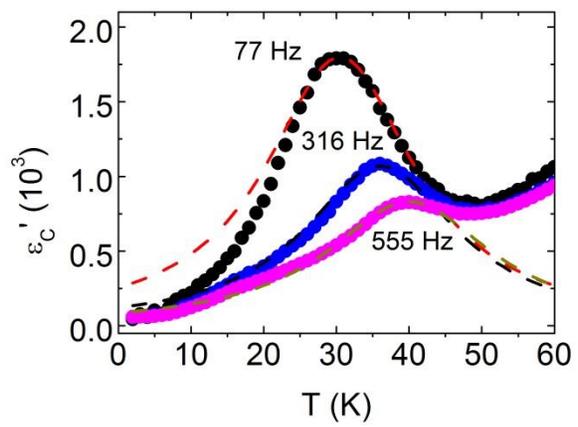

**Fig. 7**

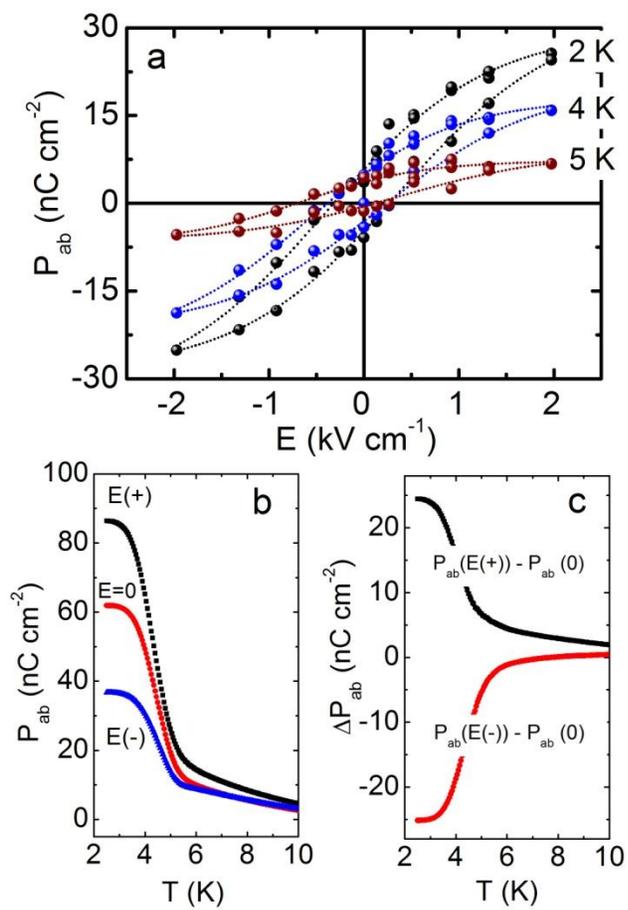

**Fig. 8**

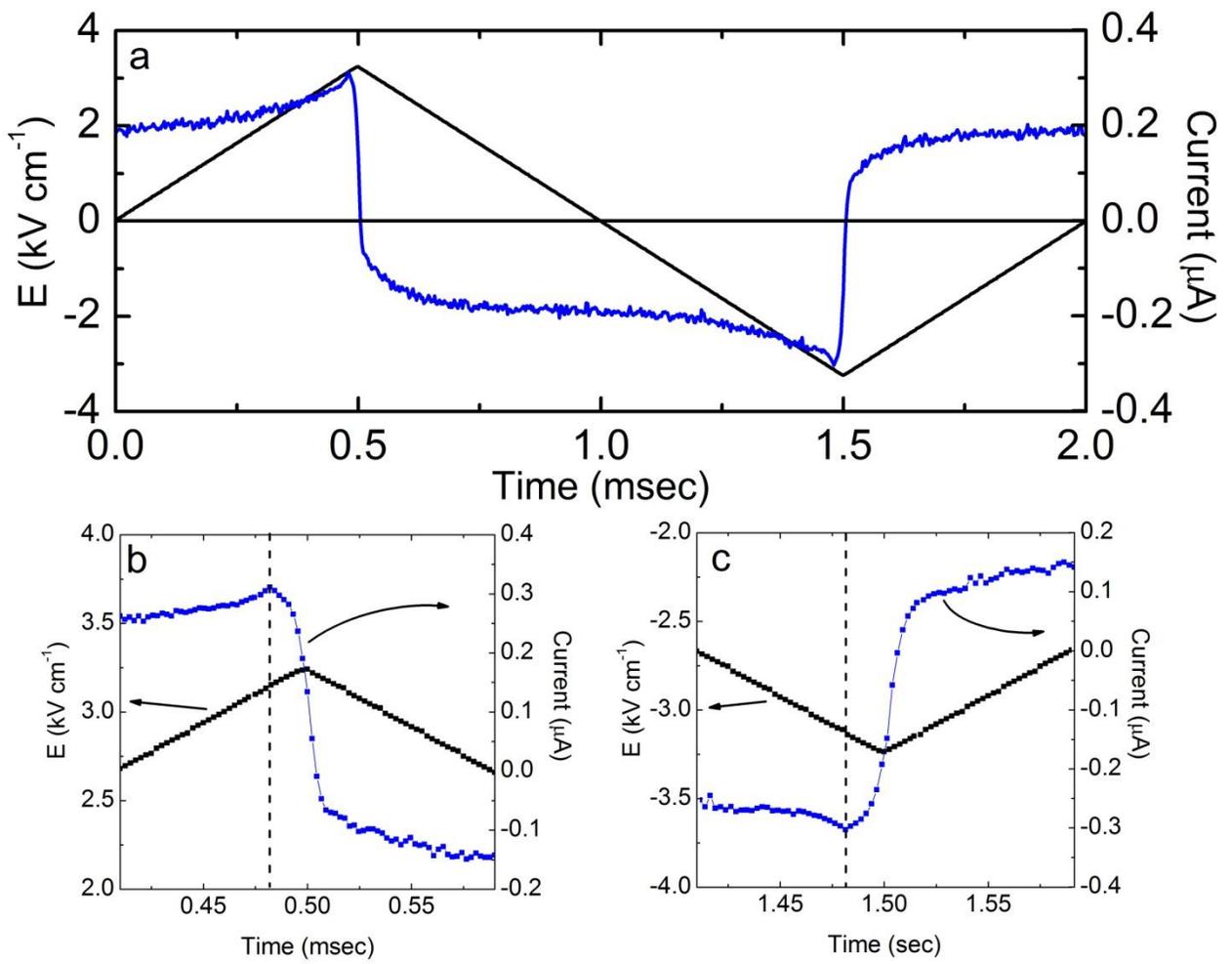

**Fig. 9**

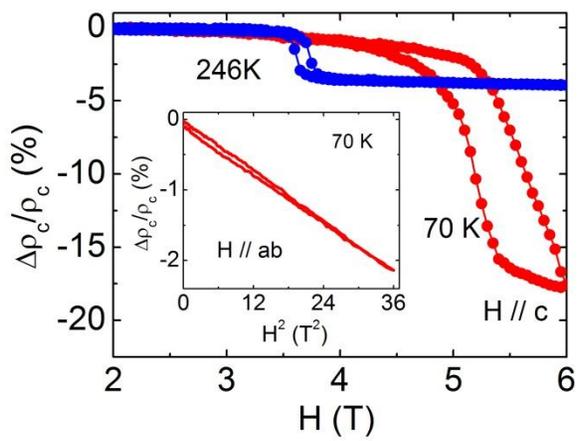

**Fig. 10**